\documentclass[]{aastex631}

\usepackage{color}

\begin{document}

\title{Deriving the interaction point between a Coronal Mass Ejection and High Speed Stream: A case study}

\author{Akshay Kumar Remeshan}
\affiliation{University of Zagreb, Faculty of Geodesy, Hvar Observatory, Ka\v{c}i\'ceva 26 10000 Zagreb, Croatia}

\author{Mateja Dumbović}
\affiliation{University of Zagreb, Faculty of Geodesy, Hvar Observatory, Ka\v{c}i\'ceva 26 10000 Zagreb, Croatia}

\author{Manuela Temmer}
\affiliation{Institute of Physics, University of Graz, Universitätsplatz 5, 8010 Graz, Austria}

\begin{abstract}
We analyze the interaction between an Interplanetary Coronal Mass Ejection (ICME) detected in situ at the L1 Lagrange point on 2016 October 12 with a trailing High-Speed Stream (HSS). We aim to estimate the region in the interplanetary (IP) space where the interaction happened/started using a combined observational-modeling approach. We use Minimum Variance Analysis and the Walen test to analyze possible reconnection exhaust at the interface of ICME and HSS. We perform a Graduated Cylindrical Shell reconstruction of the CME to estimate the geometry and source location of the CME. Finally, we use a two-step Drag Based Model (DBM) model to estimate the region in IP space where the interaction took place. 
The magnetic obstacle (MO) observed in situ shows a fairly symmetric and undisturbed structure and shows the magnetic flux, helicity, and expansion profile/speed of a typical ICME. The MVA together with the Walen test, however, confirms reconnection exhaust at the ICME-HSS boundary. Thus, in situ signatures are in favor of a scenario where the interaction is fairly recent. The trailing HSS shows a distinct velocity profile which first reaches a semi-saturated plateau with an average velocity of $500 \, \mathrm{kms}^{-1}$ and then saturates at a maximum speed of $710\, \mathrm{kms}^{-1}$. We find that the HSS interaction with the ICME is influenced only by this initial plateau. The results of the two-step DBM suggest that the ICME has started interacting with the HSS close to Earth ($\sim \, 0.81 \mathrm{AU} $), which compares well with the deductions from in situ signatures.  

\end{abstract}

\section{Introduction} \label{sec:intro}

Coronal Mass Ejections (CMEs) are violent explosive events in the solar atmosphere that result in the ejection of highly energetic magnetized plasma to interplanetary (IP) space \citep{902209}. CMEs typically consist of a compressed sheath region and a magnetic ejecta, which is presumably a flux rope \citep[e.g.,][]{Vourlidas+1997}.

Interplanetary Coronal Mass Ejection (ICME), i.e., CMEs when observed in IP space by remote sensing images or measured in situ, could cause severe space weather disturbances if Earth directed \citep[e.g.,][]{1997GMS....98...91F}. The evolution of an ICME as it traverses IP space is often a highly complex process resulting from interactions with different solar wind structures and phenomena such as high-speed streams (HSS), related stream interaction regions (SIRs), the heliospheric current sheet (HCS), or other CMEs. SIRs form by the interaction of fast solar wind from coronal holes (CHs) with the slow ambient solar wind \citep[for an overview see e.g.][]{richardson2018}. Although they are less geoeffective than ICMEs, they can also cause space weather disturbances \citep[e.g.][]{Temmer+2007,Chkhetiia+1975}. 

The interaction of CMEs with SIRs and HSSs can have multiple effects on CMEs, by altering the magnetic structure embedded in it, by introducing kinks or rotation, eroding the flux rope by magnetic reconnection, deflecting, or accelerating-decelerating it \citep[e.g.,][]{manchester}. This type of interaction may not necessarily lead to a strongly disturbed SIR, but causes significant distortions to the ICME, such as altering the $\mathrm{B_z}$ component \citep[]{duraid2018}. Moreover, numerous case studies have shown that CME-HSS interactions result in a significant increase in the magnetic field complexity \citep[e.g.,][]{Heinemann2019,mateja-cmecme2019,winslow2021,Scolini2022}. The interaction process itself may be related to reconnection signatures occurring at the interface of the ICME and HSS \citep[e.g.][]{geyer2023}. 

Among other parameters, the relative position of a CH and ICME source region, is of importance and determines where the interaction process may start, hence, how strongly either the HSS or CME magnetic field structure arriving at Earth might have changed. In this context, magnetic reconnection signatures may indicate an ongoing erosion process \citep[see e.g.,][]{ruffenach+2015}.
Additionally, it has been observed that a SIR/CIR catching up to a slower CME always tends to deform, compress, and accelerate the CME \citep[e.g.][]{efId0,winslow+2016,winslow2021,Winslow+2021b}.
The distance where CMEs and HSSs approach each other has further implications on the CME travel time \citep{manu_idist}. Knowing the distance in IP space at which the interaction occurs, and consequently, the duration of the interaction phase between the structures, is crucial for comprehending the extent of changes a CME might undergo during its transit. This will enable us to better understand and interpret in situ measurements of complex CME-SIR-HSS interaction structures at specific locations in IP space.

Since observational data is limited in IP space, modeling approaches are of great importance for studying such interaction events. Heliospheric MHD models such as the EUropean Heliospheric FORcasting Information Asset \citep[EUHFORIA;][]{EUHFORIA2018} or ENLIL \citep{odstrcil03} are convenient to analyse 3D aspects of the interaction and may be applied to derive the point of the interaction between the CME and HSS. However, the current heliospheric modeling performance is highly influenced by the unavailability of data for the photospheric magnetic field from the non-Earth-facing parts of the solar disk and our lack of understanding of the sub-Alfvénic corona \citep[see e.g., COSPAR Space Weather Roadmap update by][]{temmer2023cme}. Although these MHD models can give a good insight into the global picture; when specific time intervals and events are of interest, these can result in significant differences between actual observation and the model \citep[e.g.][]{Hinterreiter+2019}. Therefore, multiple runs would be needed to take into account uncertainty, which is computatinally very expensive. Analytical CME propagation models, such as drag-based model \citep[DBM,][]{Vrnak2013} on the other hand are very fast and can easily run huge ensembles within minutes \citep[e.g.][]{dumbovic18,calogovic21}. The application of the DBM for CME propagation and kinematics was found comparable to that of heliospheric MHD models \citep[e.g.][]{vrsnak14,yordanova24}. Therefore, if only CME kinematics is considered, and not other aspects (e.g. magnetic field), DBM provides useful means to describe CME kinematics together with its uncertainties.

In this study we present a new method to model the interaction point between the CME and HSS using DBM and analyse observations to validate the method. For this purpose we analyze a particular CME-SIR/HSS interaction. Since there is no direct way to test where the interaction occurred, we perform a detailed case study combining EUV remote sensing image data with in situ measurements, and analyse in detail in situ properties of the CME-SIR/HSS event, including a reconnection exhaust which we observe at the CME/HSS interface.

\section{Data and Method} \label{sec:datanmeth}
We study an ICME - HSS interaction event observed in situ at 1 AU on 2016-10-12 at 2200 UT. The HSS of interest is found in the CH-SIR list by \cite{Heinemann2019} and the ICME in the WIND ICME catalog from \cite{2018SoPh..293...25N}.
We use the ICME-CME association from the \citet{DVN/C2MHTH_2024} ICME list, according to which our event is related to the halo CME first appearing in LASCO C2 coronagraph's field of view at 02:24:05UT on 2016-10-09. 

We analyze in detail in situ data (see Section 2.1), remote observations (see Section 2.2), and finally use the DBM to model the CME propagation and determine the point in the heliosphere where the ICME interacts with the HSS (see Section 2.3).

\subsection{In situ data}\label{2.1}
\begin{figure*}
    \centering
    \includegraphics[scale=0.9,page=1]{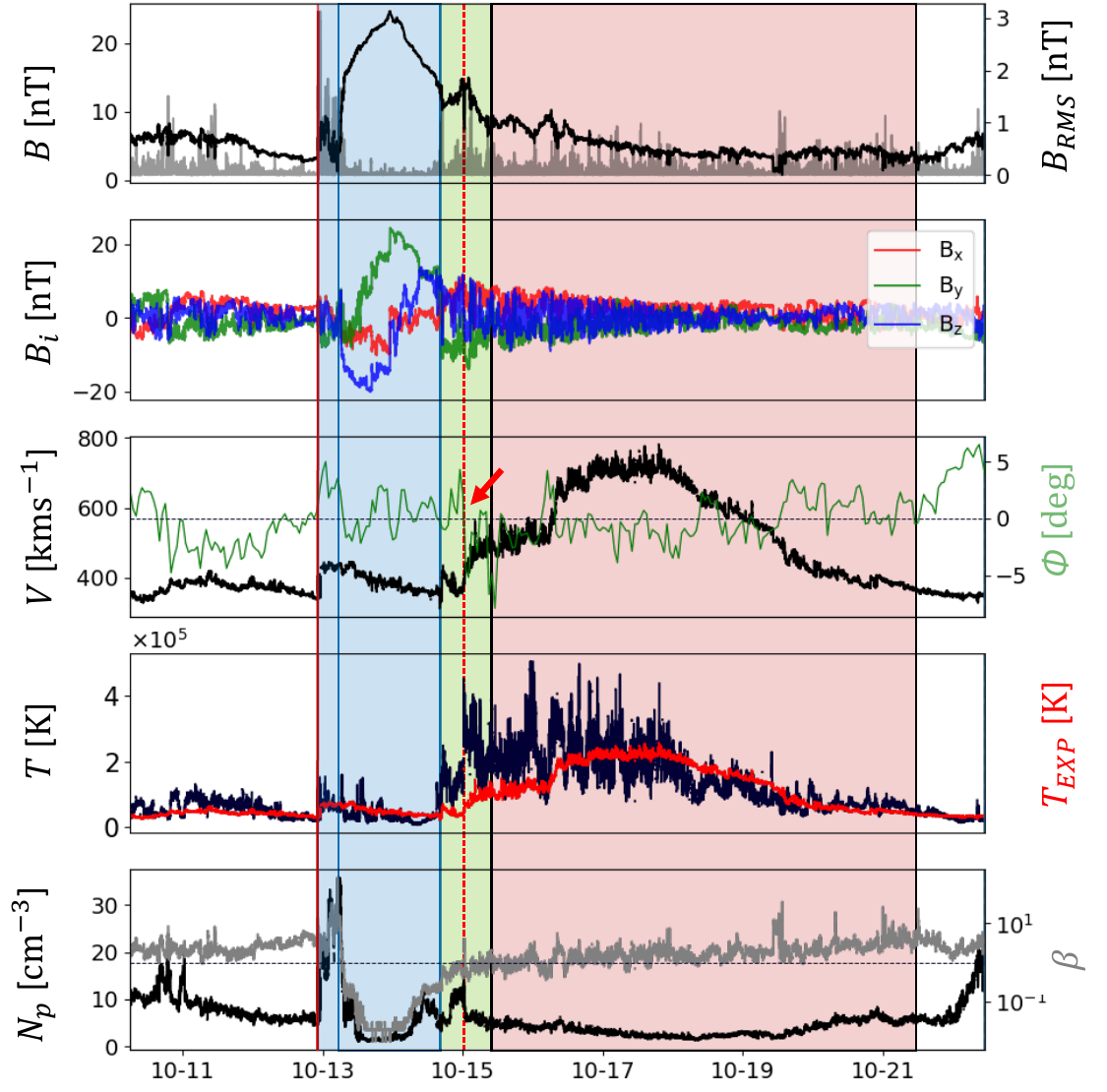}
    \caption{In situ data \textbf{for 2016-10-10 10:00 UT to 2016-10-22 00:00 UT} from OMNI database with 5-minute cadence. From top to bottom: panel 1 shows the magnetic field magnitude and variation of the magnetic field given by the root mean square (RMS); panel 2 shows the magnetic field vector components (GSE); panel 3 shows the plasma velocity; panel 4 shows the plasma proton temperature and expected plasma temperature; panel 5 shows the proton density and plasma beta (the dashed line marks $\beta =1$). The vertical red line indicates the arrival of the fast-forward shock on 2016-10-12 at 21:15UT, the vertical blue lines mark the boundaries of MO with a blue-shaded region showing the ICME including the shock-sheath region. The red region with black vertical lines as boundaries shows the HSS with an overlapping region in between them shaded in green. The vertical-dashed red line marks the stream interaction region with the red arrow pointing to the change in azimuthal flow angle. }
\label{fig:in_situ}   
\end{figure*}
To study the event in situ at 1AU, we use a combination of magnetic field and plasma data in the geocentric solar ecliptic system (GSE) from the Advanced Composition Explorer (ACE; \citep{Stone1998} and  Global Geospace Science WIND satellite \citep{1995SSRv...71....5A} data from the OMNI \citep{omni} database.

 We first analyze the large-scale properties of the event by identifying shock/sheath and the magnetic obstacle (MO) of the ICME, and HSS.

The shaded blue region in Fig. \ref{fig:in_situ} shows a sudden increase in flow velocity, temperature, and proton density on 2016-10-12 at 21:15 UT (vertical red line), marking the arrival of a fast-forward shock. These increased components are maintained together with a high $B_{\rm rms}$ till 2016-10-13 07:15:00 UT (vertical blue line), revealing a typical sheath region \citep[e.g.][and references therein]{Zurbuchen2006, Kilpua2017}.
The sheath is followed by a region of an increased total magnetic field, decreased $B_{\rm rms}$, and smooth-rotating magnetic field components (identified by a smooth change in individual magnetic field components from some maximum/minimum value to some other minimum/maximum simultaneously such that the total magnetic field magnitude is conserved), a velocity profile showing constant negative slope, and drop in plasma beta, all generic traits of an MO \citep[e.g.][]{Nieves-Chinchilla2018}.
We also investigate the profile of the expected temperature, which is an estimate for the proton temperature calculated from the proton velocity as described in \cite{1987JGR....9211189L}. A drop in the measured proton temperature below the expected temperature implies that the plasma is colder than the predicted plasma temperature for the plasma velocity in question, which is a characteristic trait of ICMEs as they expand.
These trends hold until 2016-10-14 at 16:20:00 UT, marking the end of MO.

The total duration of the ICME (recorded as the timespan from the start of the disturbance (shock) to the end of the MO is 42.8 hours, with the sheath lasting 9.4 hours and MO lasting 33 hours, which is longer by 9 hours than the average duration of MOs calculated for solar cycle 24 \citep[e.g.][]{Nieves-Chinchilla2018}. 

At the trailing edge of the ICME we observe an SIR, which is identified by noting the dip in plasma density, increase in plasma temperature, change in azimuthal flow angle, and rapid increase of velocity profile \citep{efId0}, followed by an HSS. The shaded red region in Fig. \ref{fig:in_situ}  highlights the  HSS, with the stream interface (vertical dashed red line) on 2016-10-15  at 00:30 UT. The stream interface has been set according to the change in azimuthal flow angle \citep[as in][]{efId0}.
The HSS lasts for 5 days and 12 hours, where the start of the HSS is the stream interface and the end is taken as the time when the flow speed returns to $365\,\mathrm{kms}^{-1}$ (the average solar wind speed from the day before the arrival of the ICME shock). 

We use the HSS-CH association from \cite{Heinemann2019}, which is confirmed by comparing the polarity $P$ calculated from in situ data as in \citep{2002JGRA..107.1488N}: 

\begin{eqnarray}    
P &=& \frac{B_R - \frac{\Omega R \cos \lambda B_T}{V}}{\sqrt{(1 + \left(\frac{\Omega R \cos \lambda}{V}\right)^2) \sqrt{B_R^2 + B_T^2}}}
\label{eqn:polarity}
\end{eqnarray}
 
(where $B_R\mathrm{[nT]}$, $B_T \mathrm{[nT]}$ are magnetic field components in Radial-Tangential-Normal (RTN) coordinates, $\Omega \mathrm{[S^{-1}]}$ is the solar rotation rate, $R \mathrm{[AU]}$ is the distance from the Sun and $\lambda [^{\circ}]$ is the heliographic latitude) with the polarity of the potential field source surface model (PFSS) offered in JHelioviewer \citep{2017A&A...606A..10M}.

Next, we analyze small-scale properties of the interface between the ICME and HSS, where we search for signatures of possible reconnection exhausts: A dip in the magnetic field, an increase in proton temperature, an increase in beta, and an increase in proton density \citep{https://doi.org/10.1029/2004JA010809}. 

 We transform the in situ data from GSE coordinates to the local current sheet coordinates using the hybrid Minimum Variance analysis (MVA). The transformed current sheet coordinates are such that the rotated X coordinate now lies in the plane of reconnection, the Z coordinate lies in the plane perpendicular to the current sheet, and the Y coordinate completes the right-handed orthogonal triad \citep[e.g.][]{phan+2006,wang+2010}. In the current sheet coordinate system, the following characteristic behavior must apply \citep[e.g.,][]{Mistry+2017,Tiquilin+2020}:
 \begin{itemize}
     \item Variation in $B_X$ is greater than in $B_Y$.
     \item The correlation between $V_X$ and $B_X$ changes inside the boundary.
    \item Changes in $V_X$ are greater than those in $V_Y$ and $V_Z$.
    \item The direction of $B_X$ changes inside the boundary.
    \item $B_Z$ remains close to zero at the start and end boundaries.
    
 \end{itemize}

 The start and end of the possible exhaust regions are identified by the set of criteria defined earlier and MVA is performed on a region $\pm 30 \mathrm{min}$ from the middle of the set boundaries.
The Walen test \citep[][used to check for Alfvénic exhausts that hint at magnetic reconnection]{HUDSON19701611} is then applied to confirm the presence of a magnetic reconnection event \citep[similarly as performed in ][]{geyer2023}:

\begin{eqnarray}
    V_{pred} &=& V_{ref} \pm \sqrt{\frac{1-\alpha_{ref}}{\mu_0 \rho_{ref}}} \left(\frac{\rho_{ref B}}{\rho}-B_{ref}\right)\label{eq:walen_t}
\end{eqnarray}

Where $V_{ref}$, $\alpha_{ref}$, $\rho_{ref}$ and $B_{ref}$ are the velocity, the pressure anisotropy factor, the proton density and the magnetic field, respectively, at the limits of the exhaust region. The $B$ and $\rho$ are the magnetic field and proton densities inside the boundary used to calculate the predicted velocity. We have taken the pressure anisotropy factor to be zero due to the lack of available measurements \citep[see][]{2015JGRA..120...43R}.

The $+$\textbackslash$-$ signs in the formula signify the propagation of Alfvénic waves antiparallel\textbackslash parallel to the magnetic field direction incidentally showing correlated\textbackslash anticorrelated changes between the velocity components and magnetic field components.

To quantify the magnetic properties of the MO, we then calculate the axial flux and magnetic helicity as described in \citep{devore}:

\begin{eqnarray}
    \Phi^{imc} &=& 1.4|B_0|r_0^2 
    \label{eqn: axial_flux}\\
    H^{imc}_m &=& 0.60\sigma_{H}B_0^2r_0^3l   
    \label{egn:helicity}
\end{eqnarray}

Where $\Phi^{imc} $ is the axial flux, $B_0$ the axial field, $r_0$ the radius of MO, $\sigma_H =\pm 1$ denotes the handedness of the flux rope, and $l$ the length of the flux rope. The handedness of the flux rope is determined according to the pattern of the rotation observed in magnetic field components, following \citep[e.g.][]{bothmer1998, Nieves-Chinchilla2019}. Since the CME is almost a full halo, $B_0$ is assumed to be the same value of the magnetic field at the center of the MO.
$r_0$ was determined assuming that the spacecraft passes through the center of the flux rope vertically with respect to the FR axis. $l$ is taken as $\frac{\pi}{6} \mathrm{AU}$ following calculations from \cite{devore}.

\subsection{Remote Sensing}

\begin{figure}
     \centering     
         \includegraphics[width=0.4\textwidth]{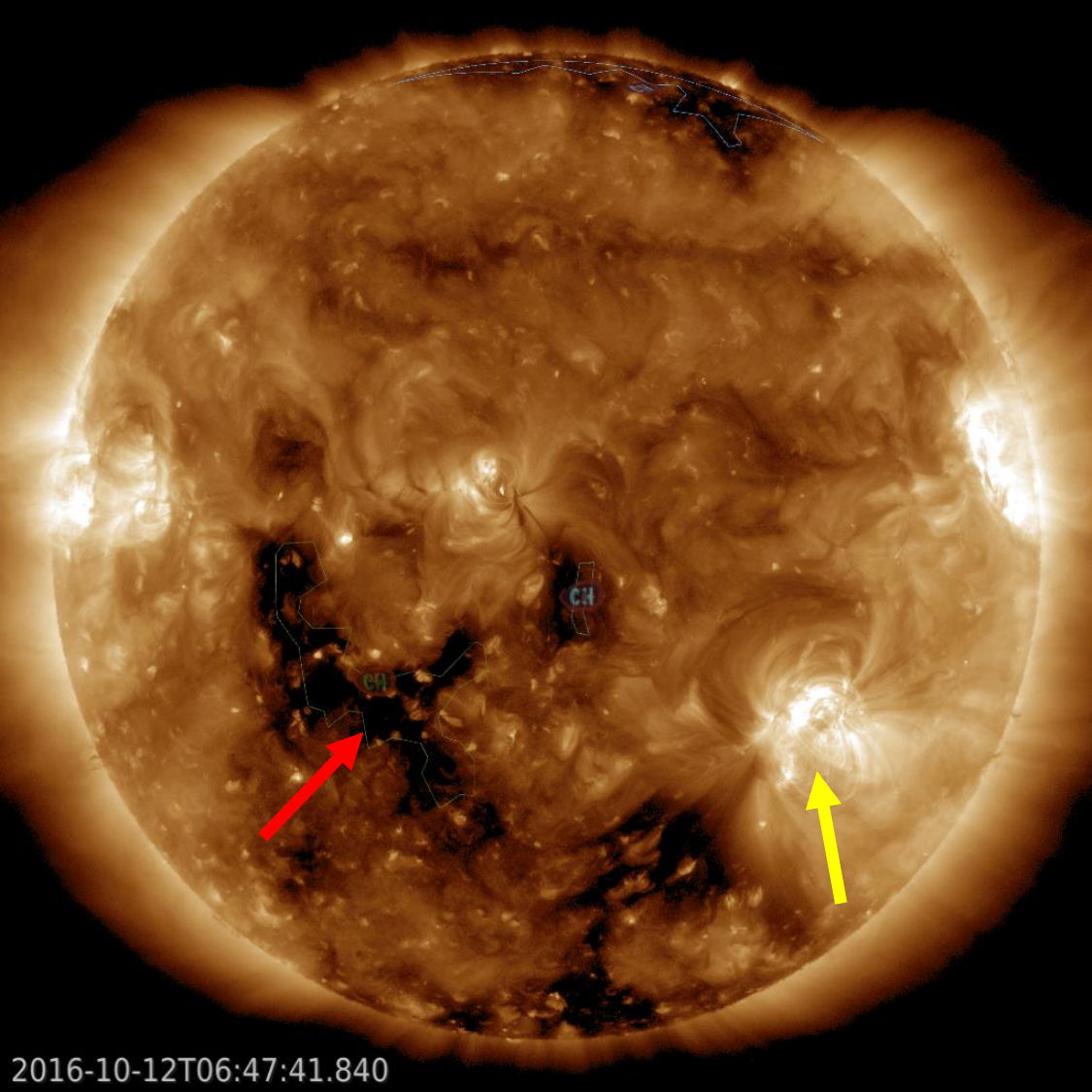}
         \caption{AIA 193{\AA} image from 2016-10-12 06:42UT showing the positions of CH (red arrow) and active region of interest (yellow arrow).}
         \label{fig:full_disk}
\end{figure}

To study the event close to the Sun, we used the C2 coronagraph from the Large Angle and Spectrometric COronagraph \citep[LASCO;][]{1995SoPh..162..357B}  onboard the SOlar and Heliospheric Observatory \citep[SOHO;][]{1995SSRv...72...81D}, COR2 coronagraph of the Sun-Earth Connection Coronal and Heliospheric Investigation \citep[SECCHI;][]{2008SSRv..136...67H} suite from the STEREO-A spacecraft, and Extreme Ultraviolet (EUV) 193\,{\AA} images from the Atmospheric Imaging Assembly \citep[AIA;][]{2012SoPh..275...17L} onboard the Solar Dynamics Observatory \citep[SDO;][]{2012AGUFMIN43E..05P} spacecraft (see Fig.~\ref{fig:full_disk}). 

We use the Graduated Cylindrical Shell \citep[GCS;][]{2011ApJS..194...33T} reconstruction to identify the source location and study the CME kinematics close to the Sun. The idealized GCS model assumes the CME to have two footpoints anchored at the solar surface,  with the CME legs formed by cones centered at the Sun and connected to a toroid with a pseudo-circular front. The whole structure is assumed to have a circular cross-section that undergoes self-similar expansion as it propagates to larger distances.
The geometry thus formed is manually refined by overlaying it on coronagraph images from different vantage points, to capture the three-dimensional geometry.

\begin{figure*}
    \centering
    \includegraphics[width=0.9\textwidth,page=1]{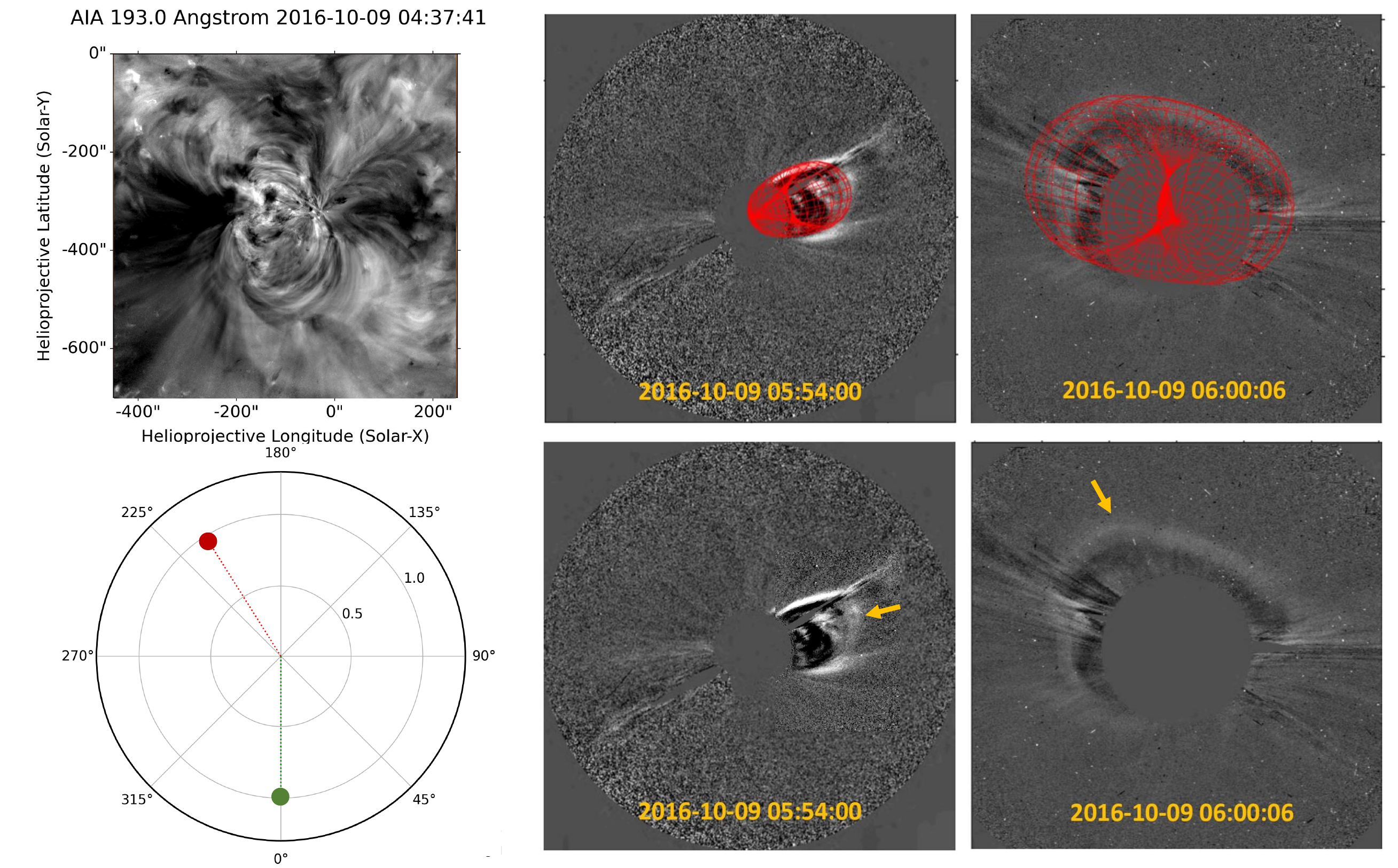}
    \caption{Top left panel shows coronal dimming visible in the cutout of the active region from SDO/AIA 193{\AA} base difference image; Four panels to the right present running difference images from STEREO-A (left) and SOHO (right) showing GCS reconstruction (top two panels) and without mesh (bottom two panels) with the CME marked with yellow arrows. The bottom left panel shows the positions of STEREO-A (red) and SOHO (green) spacecraft on 9th October 2016. The obtained GCS parameters are: Half angle = $27.5^{\circ}$, Apex height = $10\mathrm{R_{Sun}}$,  Aspect ratio $\kappa$ = 0.36, Heliographic latitude $\theta$ = $14^{\circ}$, Stonyhurst longitude $\phi$ =$355^{\circ}$,  Tilt angle = $-11.6^{\circ}$.
   }
\label{fig: GCS}
\end{figure*}

The extracted GCS parameters determine the apex center of the CME event on the solar surface with the heliographic coordinates S20E15 at 22:00UT on 2016-10-08.
We confirm the source location by identifying the associated post-eruptive coronal dimming (see the top left panel, Fig. \ref{fig: GCS}) and find that the actual source of the CME is an active region around S14E05. Appart from the error due to the subjective nature of the GCS reconstruction, this discrepancy with the extracted latitude could be due to the presence of a CH towards the southeast of the active region deflecting the CME  \citep[see, e.g.][]{2009JGRA..114.0A22G,Verbeke+2023}.

We used running difference images from the SECCHI COR2 coronagraph from STEREO-A and LASCO C2 white light coronagraph from SOHO for GCS reconstruction. 
The reconstruction was repeated several times for different time steps to obtain kinematics close to the Sun. The velocity is then estimated by performing a second-order polynomial fit.  
We also calculate the associated error in velocity using error propagation, by considering a $\pm 1\mathrm{R_{Sun}}$ error for the obtained CME-apex distance.

\subsection{Drag Based Model (DBM)}

Drag-based model \citep[DBM, ][]{Vrnak2013} is used to model the propagation of the ICME  through the heliosphere. The dynamics are calculated based on the assumption that the
Lorentz force, which initially drives the CME has a nominal effect as it leaves the corona, and motion thereafter is solely governed by the magnetohydrodynamic (MHD) drag due to interactions with the ambient solar wind and IP magnetic field. Consequently, CMEs slower than the ambient solar wind tend to get accelerated, and vice versa. The model assumes a quadratic form for the instantaneous drag acceleration given by:

\begin{eqnarray}
  a &=& -\gamma (v-w)|v-w| 
  \label{eq_dbm}  
\end{eqnarray}
    
\noindent where $v$ and $w$ are the instantaneous ICME speed and the instantaneous ambient solar wind speed, respectively and $\gamma$ is the drag parameter, which depends on the ambient solar wind density, CME mass, and cross-section of its front area. Keeping the solar wind speed and drag parameter constant, the equation can be solved analytically to give the instantaneous velocity and distance:

\begin{eqnarray}
v(t) &=& \frac{v_0-w}{1\pm \gamma (v_o-w)t}+w \label{eq_dbm1}\\
r(t) &=& \pm \frac{1}{\gamma}ln[1+\gamma(v_0-w)t] + wt +r_0 \label{eq_dbm2}
\end{eqnarray}

\noindent where $v_0$ and $r_0$ are initial CME speed and distance, respectively, $w$ is ambient solar wind speed and $\pm$ depends on the relative value of the ICME speed to the solar wind speed i.e., plus for $v_0 > w$, and minus for $v_0 < w$.

In order to take into account the CME-HSS interaction in CME propagation, we apply a two-step DBM calculation, in which the CME propagation is modeled such that it encounters the HSS in IP space. In the first step of the DBM calculation, we assume that the CME is propagating through a slow solar wind of speed $w_1$, whereas in the second step we assume that the CME is propagating in an HSS of speed $w_2$. However, a priori, we do not know at which point in the heliosphere the interaction takes place and consequently how long steps 1 and 2 should be.

Therefore, we use the following strategy to extract this information:
\begin{itemize}
    \item Step 1: Initially, the DBM with uncertainties is run from $100  $ to $ 215 \mathrm{R_{Sun}}$, assuming that the CME propagates entirely through the slow ambient solar wind.
    \item Step 2: Next, the DBM is rerun by varying the starting distances from $100  $ to $ 215 \mathrm{R_{Sun}}$. For these runs, the speed of the HSS interacting with the CME is used as the ambient solar wind speed ($w= 500 \, \mathrm{kms}^{-1}$, the average solar wind speed of the part of HSS interacting with the ICME as seen in situ; see Fig. \ref{fig:in_situ}), and the outputs from step 1 for that distance are used as inputs for the second step.
\end{itemize}

These two steps give all combinations of arrival speed, transit time, and interaction distance within the uncertainties we defined (see Fig. \ref{fig: dbm_stat}). The combinations of DBM steps that give the arrival speed matching the in situ observations will correspond to the distance at which the CME started interacting with the HSS (see Fig. \ref{fig: dbm_stat},  upper rightmost panel).

Note that there are runs which can result in the same arrival speed, but different arrival time, depending on the speed of CME ($v_{\mathrm{min}}$ or $v_{\mathrm{max}}$ of the uncertainty interval) and where the interaction starts (interaction happens earlier or later, respectively). Therefore, we use the arrival time as an additional constraint to obtain a better estimate for the interaction distance (see Fig. \ref{fig: dbm_stat},  upper middle panel).

We initialise the CME initial speed at $21.5 \mathrm{R_{sun}}$ with the value extracted from GCS reconstruction. The reconstruction is performed for multiple time steps for $ 3-20 \mathrm{R_{sun}}$, and a second order polynomial fit is used to extract the speed and associated error as $ v_0 = 357 \pm 97 \, \mathrm{kms}^{-1}$. The ambient solar wind speed $w = 387 \pm 50\,\mathrm{kms}^{-1}$, calculated as the average solar wind speed of one day before the arrival of the fast forward shock at 1AU, and the drag parameter $\gamma = 0.5 \times 10^{-7} \, \mathrm{km}^{-1}$ to initialize the model run. $\gamma$ was selected roughly according to the speed of the CME \citep[see e.g.][]{dumbovic21}.

\section{Results and Discussion} \label{sec:RnD}

\begin{figure*}[h]
    \centering
    \includegraphics[width=\textwidth]{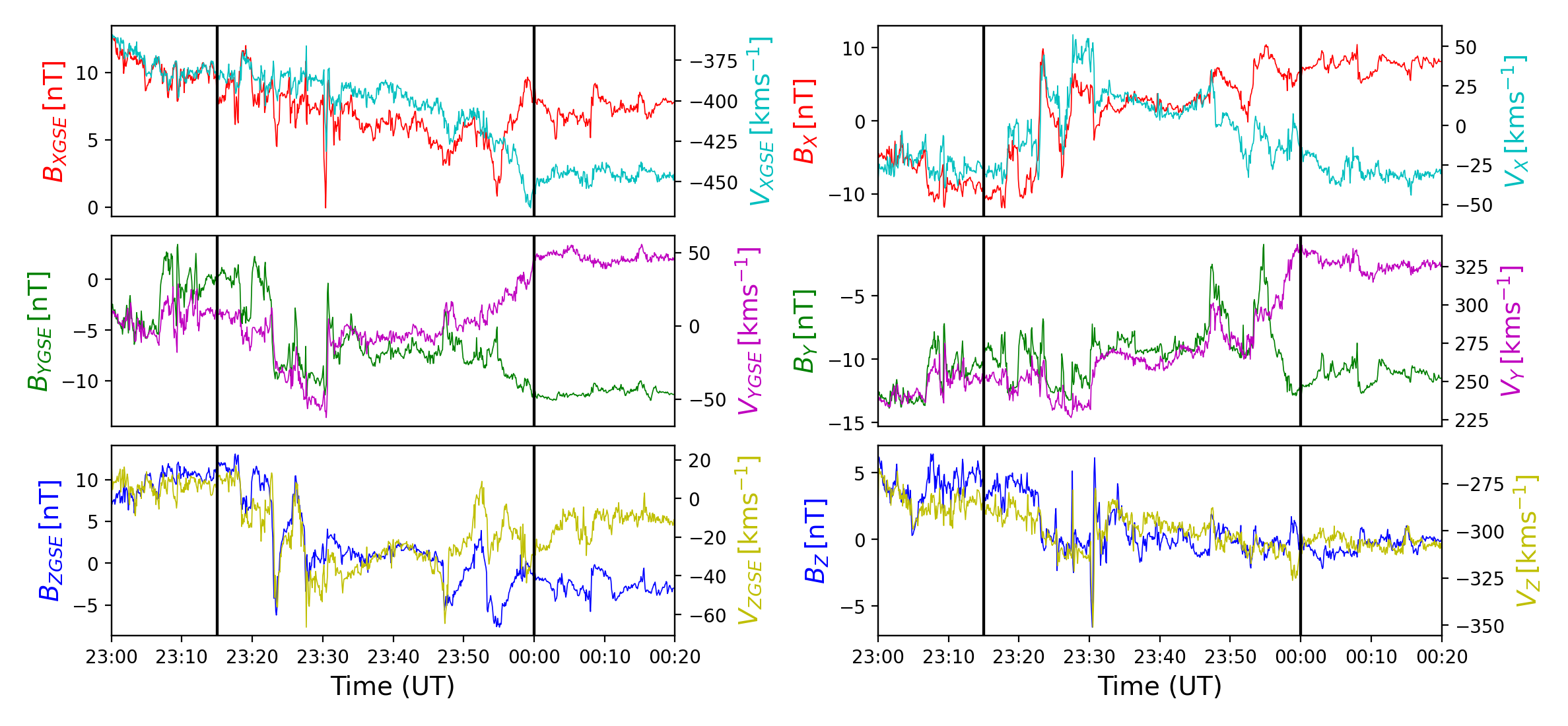}
    \caption{In situ data for 2016-10-14 23:00 UT to 2016-10-15 00:20 UT from WIND spacecraft with transformed current sheet XYZ coordinates. The left panels show the magnetic field components and plasma velocity in GSE coordinates and the right panels show the same in the transformed current sheet XYZ coordinates.}
    \label{fig:coordinate_transform}
\end{figure*}

\begin{figure}
    \centering
    \includegraphics[scale=0.32]
    {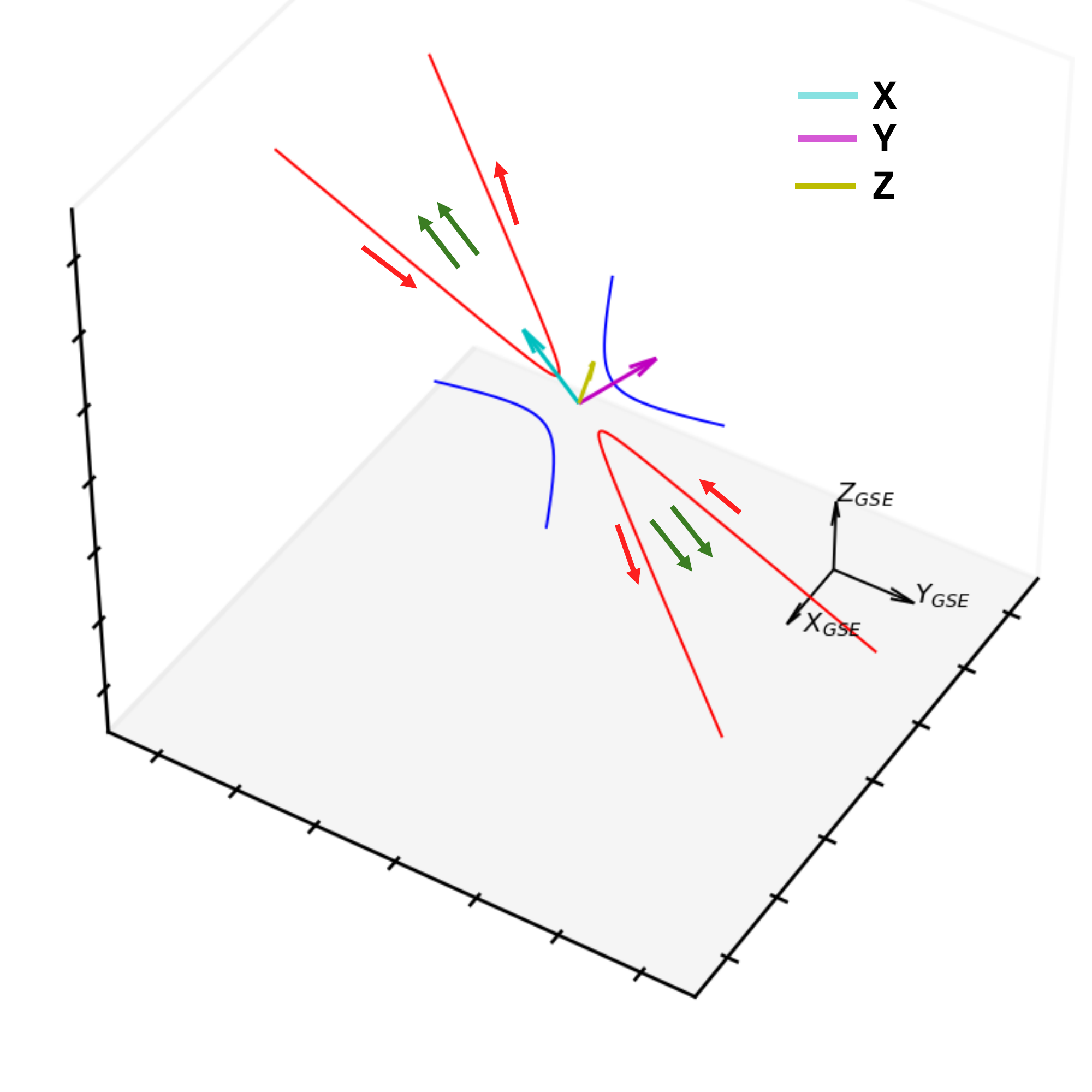}
    
    \caption{Orientation of reconnection event on the rear part of ICME in XYZ estimated using MVA with respect to GSE coordinates. Earth resides on the far side of the figure and the HSS on the near side  with the ICME between reconnection region and Earth. The red lines depict the reconnecting lines and the green arrows show the exhaust jets.}

    \label{fig: exhaust_geom}
\end{figure}

\begin{figure}
    \centering
    \includegraphics[width=0.8\textwidth,page=1]{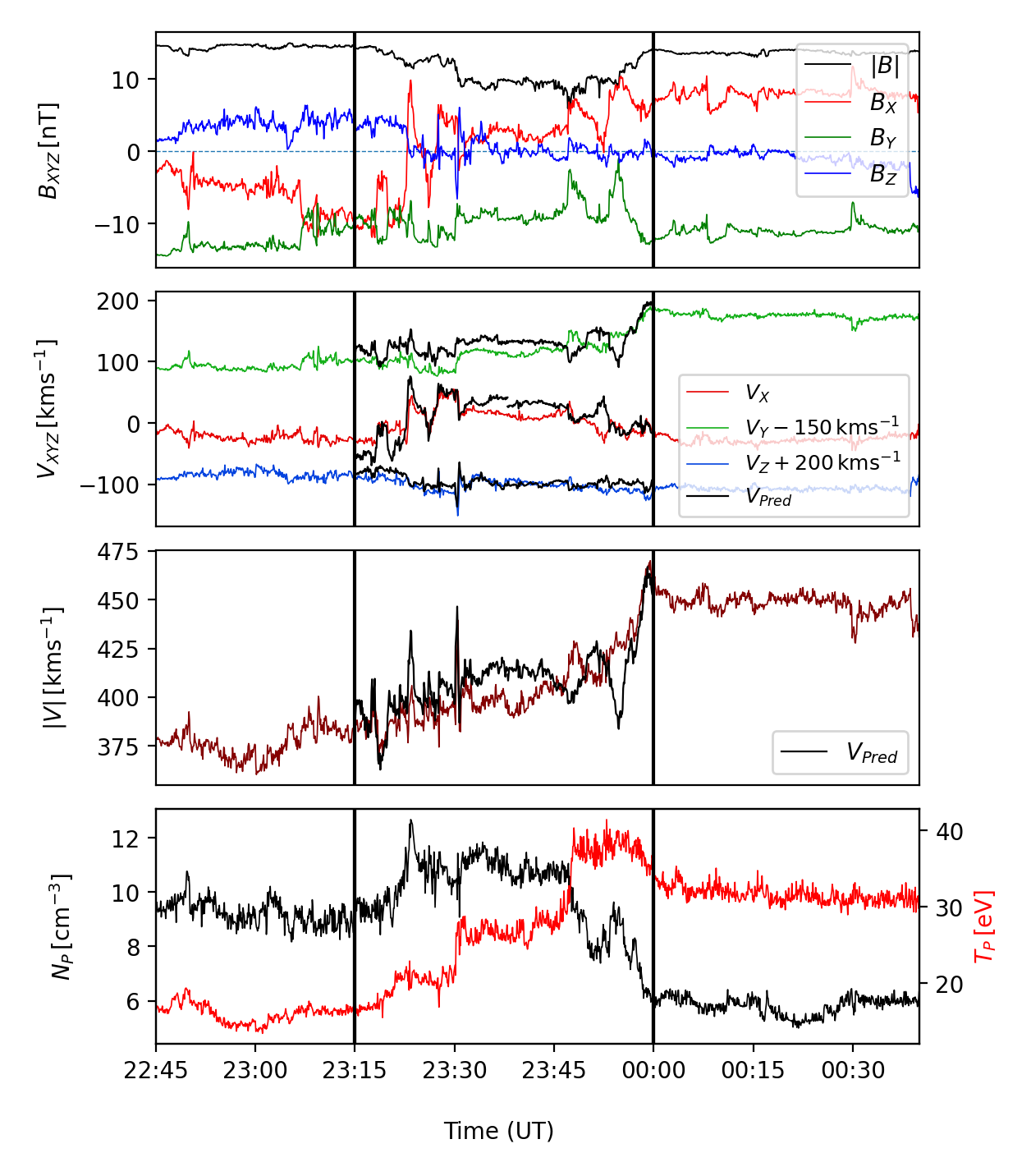}
    \caption{In situ data for 2016-10-14 22:45 UT to 2016-10-15 00:45 UT from WIND spacecraft showing reconnection exhaust region with Walen test results. From the top, panel 1 shows the magnetic field magnitude and components in the transformed current sheet coordinates. Panel 2 shows the velocity components in the transformed current sheet coordinates. Panel 3 shows the magnitude of plasma velocity and panel 4 shows plasma density and temperature. The vertical black lines mark the boundaries of the exhaust region (corrected for time shift in the OMNI database) with the solid black lines in panel 2 and panel 3 showing the corresponding predicted plasma velocities.}
\label{fig: walen}  
\end{figure}  

\begin{figure*}
\centering
\includegraphics[width=\textwidth]{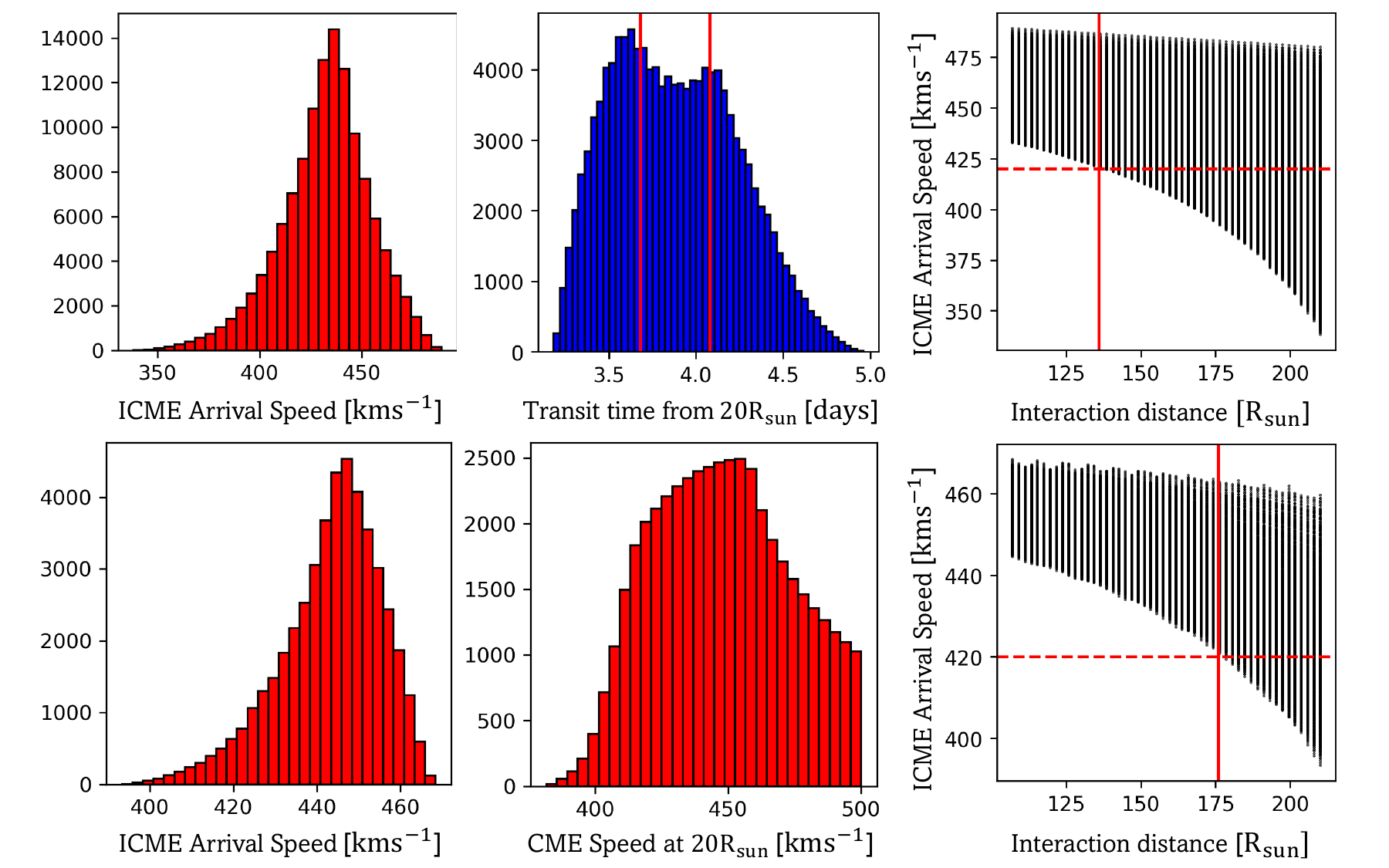}
\caption{The top left panel shows the statistics of the arrival speed from two-step DBM with a mean of $434\, \mathrm{kms}^{-1}$. The top middle panel shows the statistics for transit time in days with a mean of $3.8\, \mathrm{days}$. The top right panel shows the possible combinations of CME $v_0$ and solar wind $w$ for the respective combination of arrival speed and interaction distance from the Sun. 
The bottom left panel shows the statistics of the arrival speed from two-step DBM after constraining transit time to $\pm \, 5hrs$ from the observed in situ arrival time. The bottom middle panel shows the plausible initial speed for two-step DBM after constraining transit time with a mean of $447\, \mathrm{kms}^{-1}$. The bottom right panel shows the possible combinations of ICME $v_0$ and solar wind $w$ for the respective combination arrival speed and interaction distance from the Sun.
The red vertical lines in the top middle panel mark the boundary for the observed transit time with an added uncertainty of $\pm$5 hours.
The vertical red line of the bottom panel marks the interaction point for which at least one combination of ICME $v_0$ and solar wind $w$ results in the observed arrival speed (horizontal-dashed red line) in situ. The bottom right panels show the same, but for constrained measurements}
\label{fig: dbm_stat}
\end{figure*}

We first analyze the global properties of the ICME. Figure \ref{fig:in_situ} shows the combined in situ observations of the ICME-HSS interaction event that we studied. The observed in situ data show the characteristics of a typical ICME (See sect. \ref{2.1}) marked as the shaded blue region. Panel 2 shows a velocity profile with a clear self-similar expansion signature for the MO with a calculated radial expansion velocity of $40\,\mathrm{kms}^{-1}$ (9.5 \% of the ICME propagation speed), which compares well to the average ICME expansion speed \citep[see e.g.][]{Nieves-Chinchilla2018}. 

The MO shows a symmetric flux rope structure with its $\mathrm{B_z}$ component first reaching negative values and then changing to positive around the middle point of the MO time frame. The $\mathrm{B_y}$ component stays positive and gains maximum value around the same point, and the $\mathrm{B_x}$ component stays almost constant. The behavior of the magnetic field components points to a flux rope with a South-East-North Left-handed (SEN-LH) chirality with a slight negative tilt \citep[according to classification in ][]{bothmer1998}. We note that the tilt agrees well with our GCS reconstruction. The magnetic field components show a rotation larger than $180^{\circ}$ (F+ flux rope structure according to \citet{Nieves-Chinchilla2019} with the total magnetic field reaching 24.2nT on 2016-10-13 around 22:35:00 UT. We also note that according to \citet{Nieves-Chinchilla2019}, the magnetic field fitting results for this event are equally good for both the symmetric circular cross-section model and the elliptical cross-section model.

To calculate the axial magnetic flux and helicity of the FR we use $B_0=24.2 \mathrm{nT}$ (taken as the magnitude of $B$ at the center of the MO time frame), $r_0=0.153 \mathrm{AU}$ \citep[taken from the circular cylindrical flux rope model fit on in situ data from][]{Nieves-Chinchilla2019} and $l =\pi/6 \mathrm{AU}$ \citep[taken as the flux rope length calculated for a CME with half angle $30^{\circ}$ from][]{devore}. These are inputs for Equations \ref{egn:helicity} (with appropriate unit conversion), which produce the result for the axial flux ($\Phi^{imc}$) and magnetic helicity $H_m^{imc}$ as $1.9\times 10^{22} \mathrm{Mx}$ and $4.1 \times 10^{42}\mathrm{ Mx}^2$ respectively. These are typical expected values for the IP flux rope \citep[e.g.,][]{devore}, indicating that no large erosion has occurred.

Based on our in situ analysis, we conclude that the ICME shows typical behavior without prominent signs of distortion due to interaction with the HSS. This might indicate that the interaction is a relatively "young" process in the ICME evolution, hence, started at a distance rather close to 1AU. 

Next, we focus our analysis on the region just after the end of MO (shaded green in Fig. \ref{fig:in_situ}), which shows the area of overlap between the ICME and the HSS. This part shows a slightly increased $B_{\rm rms}$ and a spike in plasma beta. We also observed a dip in the magnetic field, a spike in proton density, an increase in proton temperature, and an increase in velocity between 2016-10-15 00:05 UT and 2016-10-15 00:55 UT, suggesting a possible reconnection exhaust \citep{https://doi.org/10.1029/2004JA010809}.

 The right panels in Fig.\ref{fig:coordinate_transform} show the transformed current sheet coordinates of the magnetic field and velocity components. We see that between 2016-10-15 23:15 UT and 2016-10-15 00:00 UT, the $B_X$ components change from around $-10 \mathrm{nT}$ to $+9 \mathrm{nT}$ while the $B_Y$ component is relatively unchanged and stays around $-10 \mathrm{nT}$. Similarly, the value of $B_Z$ stays around $0 \mathrm{nT}$. Looking at the velocity components in the same interval, we see that the variation in $V_X$ is the largest, and the correlation between the velocity components changes towards the end of the interval.   

The performed hybrid MVA analysis thus reveals that the required criteria for a reconnection exhaust region match within some error bars.
The geometry of the revealed exhaust system is shown in Fig. \ref{fig: exhaust_geom}.

Figure \ref{fig: walen} shows the result of the Walen test performed between  2016-10-14 23:15 UT and 2016-10-15 00:00 UT. We find that the front region shows correlated changes between magnetic field components, and the back region shows anti-correlated changes between the same. The bottom panel from Fig. \ref{fig: walen} also shows increased plasma density (black) and increased plasma temperature (red). The Walen test together with MVA and hints of increased plasma temperature, plasma density and plasma beta hence confirms the existence of a reconnection exhaust at the interface between ICME and HSS.
We note that the region of interest shows a rather long temporal width ($\mathrm{40min}$) compared to other studies that point to an average width of a few minutes \citep[e.g.,][]{Mistry+2017}.

In the total velocity profile of typical reconnection exhausts, a small bell-like increase can be observed, which is a consequence of the outflow observed in the x-direction of the current sheath coordinate system. We note that the outflow, which is observed in the x-direction, is not reflected in the total velocity component. This is because the change from slow CME to fast HSS, which is in radial direction is a dominant feature compared to the small bump corresponding to the outflow, which is in non-radial direction. This is thus, another specificity of this reconnection exhaust.

The trailing HSS has a negative polarity and a velocity profile showing a two-part rise. The flow velocity increases after the HSS-ICME interface and reaches a semi-saturated plateau (showing a slight increase over time) that lasts around 28 hours with an average velocity of $500\, \mathrm{kms}^{-1}$ before again increasing to a maximum value of $710\, \mathrm{kms}^{-1}$. This behavior could be due to the patchy nature of the source CH of the HSS, which also agrees with remote observation (see Fig. \ref{fig:full_disk}).

Figure~\ref{fig: dbm_stat} shows the results of an exercise using two-step DBM runs to identify the ICME-HSS interaction distance. Using solely the values from remote observations near the sun, we find the ICME mean arrival speed to be $434\, \mathrm{kms}^{-1}$ with a standard deviation of $21.8\,\mathrm{kms}^{-1}$
and a mean transit time of $3.8\,\mathrm{days}$ with a standard deviation of $0.36\,\mathrm{days}$. The DBM results for ICME arrival velocity are within $3.2\%$ of the measured in situ velocity at the start of the MO, and the ICME transit time (i.e., the ICME propagation time from 0.1\,AU to 1AU) is within $2.3\%$ of the measured transit time.

The top middle panel of Fig. \ref{fig: dbm_stat} shows two subtle yet distinct peaks for the arrival time. This trend is possibly due to the different combinations of solar wind speed and CME speed at each interaction point that could result in similar arrival speeds but distinct transit times. We can see this from the top right panel of Fig. \ref{fig: dbm_stat}, where each vertical line in the figure represents the combinations of two-step DBM parameters that result in the corresponding arrival speed of the ICME. This result suggests possible combinations of $v_0$ and $w$ for interaction points greater than $136\, \mathrm{R_{Sun}}$ that can result in the same in situ measured mean arrival speed of $\mathrm{420\, kms^{-1}}$. This result suggests a plausible interaction region as large as 0.4AU (130 to 215 $\mathrm{R_{Sun}}$)!

\begin{figure}
    \centering
    \includegraphics[scale=0.24,page=3
]{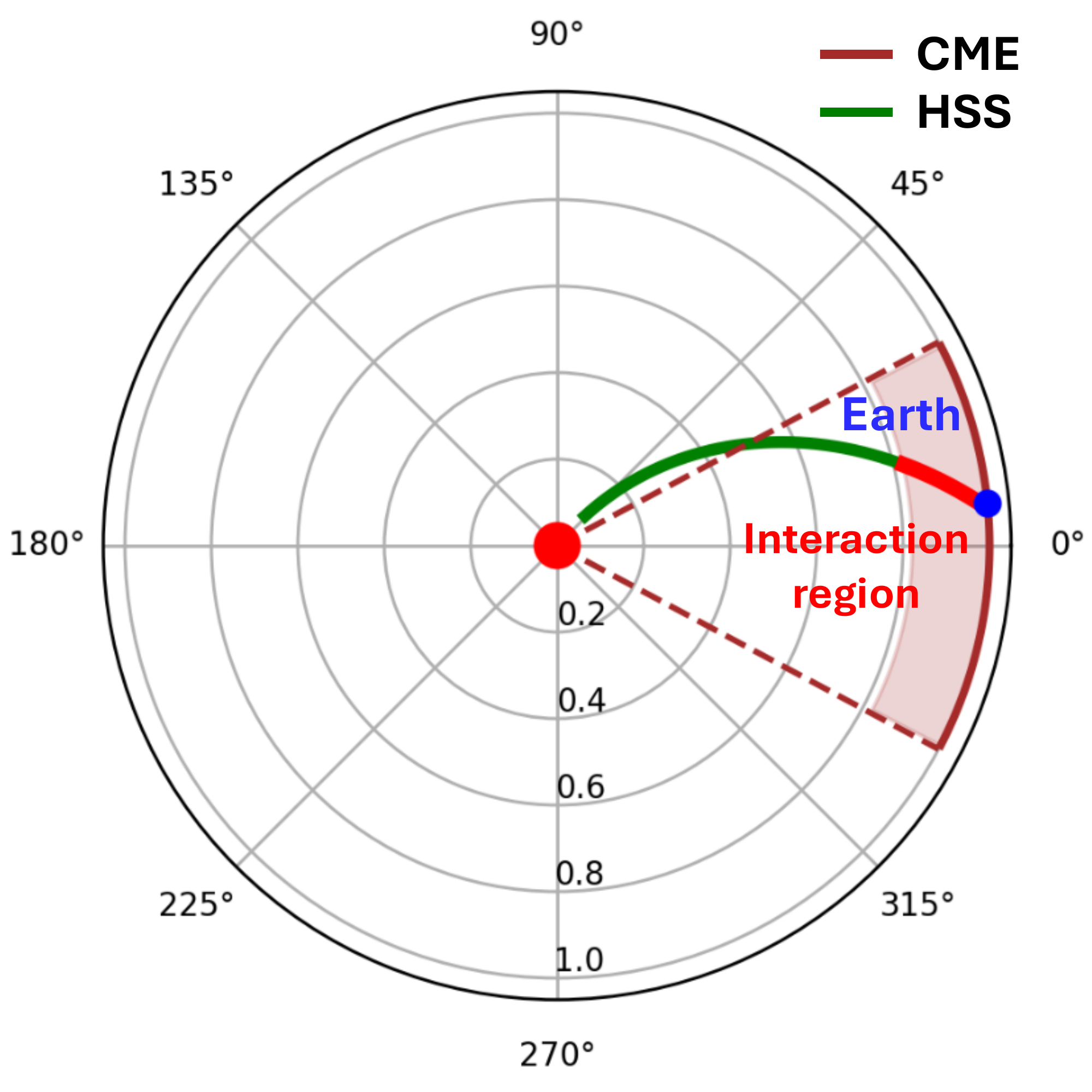}
    \caption{ICME-HSS interaction region estimated from the two-step DBM. The brown cone shows the CME opening angle. The green curved line shows the HSS and The region shaded red region shows the interaction region
.}
    \label{fig:interacion_fig}
\end{figure}

We therefore constrain the transit time to a reasonable $\pm \, 5 \mathrm{h}$ from the observed 3.78 days. The bottom panels in Fig. \ref{fig: dbm_stat} give the results after applying the constraint to the transit time. The bottom left panel shows that the mean arrival speed has changed to $443\,\mathrm{kms}^{-1}$ and the standard deviation has been reduced to $12\,\mathrm{kms}^{-1}$.
From the bottom right panel, we see that the possible interaction region is now above $175\, \mathrm{R_{Sun}}$  (0.81AU). The estimate of interaction distance from the constrained two-step DBM now compares well with our deduction from in situ analysis (see Fig.~\ref{fig:interacion_fig}). In addition, we see from the bottom middle panel that the CME initial velocity now has an estimate with a mean of $447\,\mathrm{kms}^{-1}$ and a standard deviation of $27\,\mathrm{kms}^{-1}$. The values are well within the estimated range of the CME initial velocity $356\,\pm 97\, \mathrm{kms}^{-1}$ derived from the kinematics obtained from the GCS reconstruction.

To validate the two-step DBM results, we find that using the initial conditions solely from remote observations near the Sun results in a reasonably good prediction of arrival speed for the CME. However, this approach gives an unrealistic estimate of the total transit time and interaction distance. The result is physically meaningful as an early interaction between ICME and HSS can accelerate the ICME from the lower bound of the initial velocity range, which incidentally results in an altered transit time. Similarly, a later interaction can mean that the initial velocity from the upper bound of the initial velocity range could result in the same arrival speed and altered transit time. 
Thus, constraining the transit time from the observations enables us to filter out this effect, and the resulting statistics for initial velocity naturally change the initial velocity distribution from the uniform distribution (our first assumption), reducing the standard deviation and giving a comparable estimate to that obtained from GCS reconstruction. We also note that adding a constraint for the arrival speed of $ 420 \pm 20\,\mathrm{kms}^{-1}$ did not have any notable effect on plausible combinations for the initial parameters or the interaction distance. This observation supports our assumption of having multiple combinations of initial parameters that could result in the same arrival speed but distinct transit times. This inference therefore signifies the importance of using arrival time as a constraint to negate the effects of multiple combinations of interaction distances and initial speed. 

\section{Summary and Conclusion}

We analyzed the ICME observed in situ on 2016-10-12 at 1AU with a trailing HSS interacting with it. We associate the ICME and HSS with their solar sources, using existing catalogs and confirming it via comparison of remote and in situ observation.

The in situ data show a typical ICME, with a symmetric magnetic field structure, while the plasma velocity shows an unhindered expansion profile. The symmetric structure of the flux rope compares well with the halo nature of the associated CME seen in the LASCO C2 coronagraph and further suggests that the in situ measurements were taken around the apex of the ICME, with the spacecraft passing near the center of the flux rope. As the ICME shows no prominent signs of distortion, this might hint towards only a weak - just recently started - or no interaction with the following HSS.

However, a more detailed analysis of the interface between ICME and HSS does reveal a magnetic reconnection exhaust. Given that a lenience is kept in mind knowing wider regions can show larger fluctuations as it has more possibilities to get disturbed by ambient conditions, although temporally much wider compared to other studies, we see that the criterion set to establish the "genuineness" is satisfied at this region.
This clearly indicates an ongoing interaction process between ICME and HSS. 
If the interaction, hence reconnection, would be already long-lasting, it possibly would have significantly eroded the ICME, and that might have distorted the ICME signatures.

This recent interaction is found to occur at the "first-plateau" region, observed in the HSS velocity profile with an average velocity of $500\,\mathrm{kms}^{-1}$. Results from the two-step DBM support well this inference. The result of the initial DBM run covering all possible combinations of initial parameters within the chosen range showed plausible results for the measured arrival time and speed (with high confidence intervals) within a large interaction distance range of 130 -- 215\, $\mathrm{R_{Sun}}$. But constraining the time of transit to a reasonable $\pm \, 5 \mathrm{h}$ to the calculated transit time improves the estimate of arrival speed reducing the standard deviation from 21 to 12\,$\mathrm{kms}^{-1}$. This step also pushes the result for the most plausible interaction region to 0.81AU (175\,$\mathrm{R_{Sun}}$), which agrees well with our conclusion based on in situ measurements.

To conclude, this case study demonstrates how a two-step DBM approach can be successfully applied to determine the point of interaction between an ICME and HSS. The event we have analyzed covers an interaction very close to $1\, AU$, hence, the complexity in the in situ ICME-HSS signatures resulting from such a recent interaction is found to be low and the ICME flux rope part stays rather intact. Due to the low computational cost, the analytical DBM model is well suited to investigate ICME-HSS interaction events over various distances in a statistical manner. This will give further insight into the distortion of flux ropes from HSS interactions and how that affects ICME transit times and arrival speeds.

\section*{Acknowledgments}
We acknowledge the support by the Croatian Science Foundation under the project IP-2020-02-9893 (ICOHOSS) and from the Austrian-Croatian Bilateral Scientific Project ”Analysis of solar eruptive phenomena from cradle to grave”. A.K.M. acknowledges support from the Croatian Science Foundation in the scope of Young Researches Career Development Project Training New Doctoral Students. We thank the science teams of SDO, SOHO, STEREO, ACE, and Wind for providing the data and maintaining the spacecraft. This study has made use of the JHelioviewer software provided by ESA.

\vspace{5mm}
\bibliography{main}{}
\bibliographystyle{aasjournal}
\end{document}